\RequirePackage{fix-cm}
\documentclass[smallextended]{svjour3}       
\smartqed  
\usepackage{booktabs} 
\usepackage[ruled]{algorithm2e} 
\usepackage{graphicx}
\usepackage{color, soul, xcolor}  
\usepackage{colortbl}
\usepackage{booktabs} 
\usepackage{enumitem}
\usepackage[normalem]{ulem}
\usepackage{caption}
\usepackage[caption=false,font=footnotesize,labelformat=simple]{subfig}
\usepackage{amsmath}
\usepackage{makecell}
\usepackage{balance}
\usepackage{framed}
\definecolor{shadecolor}{gray}{0.95}
\usepackage[round]{natbib}   

\begin{document}


\title{Bugs in Infrastructure as Code}
\titlerunning{Bugs in Infrastructure as Code}        

\author{Akond Rahman          \and
        Sarah Elder           \and 
        Faysal Hossain Shezan \and  
        Vanessa Frost         \and 
        Jonathan Stallings    \and 
        Laurie Williams
}


\institute{Akond Rahman \at
              Department of Computer Science, North Carolina State University, Raleigh, NC, USA \\
              \email{aarahman@ncsu.edu}           
           \and
           Sarah Elder \at
              Department of Computer Science, North Carolina State University, Raleigh, NC, USA  \\
              \email{seelder@ncsu.edu}
           \and
           Faysal Hossain Shezan \at
              Department of Computer Science, University of Virginia, Charlottsville, VA, USA \\
              \email{faysalhossain2007@gmail.com}
           \and
           Vanessa Frost \at
              Department of Computer Science, University of Florida, Gainesville, FL, USA \\
              \email{nyxciardha@gmail.com}
           \and
           Jonathan Stallings \at
              Department of Statistics, North Carolina State University, Raleigh, NC, USA  \\
              \email{jwstalli@ncsu.edu}
           \and
           Laurie Williams \at
              Department of Computer Science, North Carolina State University, Raleigh, NC, USA  \\
              \email{williams@cs.ncsu.edu}
}

\date{Received: date / Accepted: date}

\maketitle


\begin{abstract}

Infrastructure as code (IaC) scripts are used to automate the maintenance and configuration of software development and deployment infrastructure. IaC scripts can be complex in nature and can contain hundreds of lines of code, leading to defects that can be difficult to debug and to wide-scale system discrepancies, such as service outages. The use of IaC scripts is getting increasingly popular, yet the nature of defects that occur in these scripts have not been systematically categorized. \textit{The goal of this paper is to help software practitioners improve their development process of infrastructure as code (IaC) scripts by analyzing the defect categories in IaC scripts based upon a qualitative analysis of commit messages and issue report descriptions.} We collect 12,875 commits that map to 2,424 IaC scripts from four organizations, namely Mirantis, Mozilla, Openstack, and Wikimedia Commons. With 89 raters, we apply the defect type attribute of the orthogonal defect classification (ODC) methodology to categorize the defects. We also review prior literature that has used ODC to categorize defects in non-IaC software systems, and compare the defect category distribution of IaC scripts with 26 non-IaC software systems. From our analysis, we observe the dominant defect category to be `assignment', which includes defects related to syntax and configuration errors. Accordingly, the ODC process improvement guidelines recommend the teams allocate more code inspection, static analysis, and unit testing effort for IaC scripts. We also observe defects categorized as assignment to be more prevalent amongst IaC scripts compared to the 26 non-IaC software systems.


\end{abstract}


\section{Introduction}
\label{intro}

Continuous deployment (CD) is the process of rapidly deploying software or services automatically to end-users~\citep{me:agile:cd2015}. The use of infrastructure as code (IaC) scripts is essential to implement an automated deployment pipeline, which facilitates continuous deployment~\citep{Humble:2010:CD}. Companies such as Netflix~\footnote{https://www.netflix.com/}, Ambit Energy~\footnote{https://www.ambitenergy.com/}, and Wikimedia Commons~\footnote{https://commons.wikimedia.org/wiki/Main\_Page}, use IaC scripts to automatically manage their software dependencies and construct automated deployment pipelines~\citep{parnin:adages}~\citep{ambit:pup}. The use of IaC scripts helps organizations to increase their deployment frequency. For example, Ambit Energy uses IaC scripts to increase their deployment frequency by a factor of 1,200~\citep{ambit:pup}.

Defects in IaC scripts can have serious consequences as companies, such as Wikimedia, use these scripts to provision their servers and ensure availability of services~\footnote{https://blog.wikimedia.org/2011/09/19/ever-wondered-how-the-wikimedia-servers-are-configured/}. IaC scripts are susceptible to human errors~\citep{parnin:adages} and bad coding practices~\citep{cito:fse2015:cloud}, which make scripts susceptible to defects~\citep{JiangAdamsMSR2015}~\citep{parnin:adages}. Any defect in a script can propagate at scale, leading to wide-scale service discrepencies. For example in January 2017, execution of a defective IaC script erased home directories of around 270 users in cloud instances maintained by Wikimedia~\footnote{https://wikitech.wikimedia.org/wiki/Incident\_documentation/20170118-Labs}. Prior research studies~\citep{JiangAdamsMSR2015}~\citep{parnin:adages} and the above-mentioned Wikimedia incident motivate us to systematically study the defects that occur in IaC scripts.     

Categorization of defects can guide organizations on how to improve their development process. Chillarege et al.~\citep{odc:original} proposed the orthogonal defect classification (ODC) technique which included a set of defect categories. According to Chillarege et al.~\citep{odc:original}, each of these defect categories map to a certain activity of the development process which can be improved. Since the introduction of ODC in 1992, companies such as IBM~\citep{butcher:ibm:odc}, Cisco~\footnote{https://www.stickyminds.com/sites/default/files/presentation/file/2013/t12.pdf}, and Comverse~\citep{comverse:odc:ram} have successfully used ODC to categorize defects. Such categorization of defects help practitioners to identify process improvement opportunities for software development. For example, upon adoption of ODC practitioners from IBM~\citep{butcher:ibm:odc} reported ``\textit{The teams have been able to look at their own data objectively and quickly identify actions to improve their processes and ultimately their product. The actions taken have been reasonable, not requiring huge investments in time, money, or effort}". A systematic categorization of defects in IaC scripts can help in understanding the nature of IaC defects and in providing actionable recommendations for practitioners to mitigate defects in IaC scripts. 

  
\textit{The goal of this paper is to help practitioners improve their development process of infrastructure as code (IaC) scripts by categorizing the defects in IaC scripts based upon a qualitative analysis of commit messages and issue report descriptions.}

We investigate the following research questions:  

\begin{description}[leftmargin=*]
\item{\textbf{\textit{RQ1: How frequently do defects occur in infrastructure as code scripts?}}}
\item{\textbf{\textit{RQ2: By categorizing defects using the defect type attribute of orthogonal defect classification, what process improvement recommendations can we make for infrastructure as code development?}}}
\item{\textbf{\textit{RQ3: What are the differences between infrastructure as code (IaC) and non-IaC software process improvement activities, as determined by their defect category distribution reported in the literature?}}}
\end{description}

We use open source datasets from four organizations, Mirantis, Mozilla, Openstack, and Wikimedia Commons, to answer the three research questions. We collect 20, 2, 61, and 11 repositories respectively from Mirantis, Mozilla, Openstack, and Wikimedia Commons. We use 1021, 3074, 7808, and 972 commits that map to 165, 580, 1383, and 296 IaC scripts, collected from Mirantis, Mozilla, Openstack, and Wikimedia Commons, respectively. With the help of 89 raters, we apply qualitative analysis to categorize the defects that occur in the IaC scripts using the defect type attribute of ODC~\citep{odc:original}. We compare the distribution of defect categories found in the IaC scripts with the categories for 26 non-IaC software systems, as reported in prior research studies, which used the defect type attribute of ODC and were collected from IEEEXplore~\footnote{https://ieeexplore.ieee.org/Xplore/home.jsp}, ACM Digital Library~\footnote{https://dl.acm.org/}, ScienceDirect~\footnote{https://www.sciencedirect.com/}, and SpringerLink~\footnote{https://link.springer.com/}.
 
We list our contributions as following: 
\begin{itemize}[leftmargin=*]
\item{A categorization of defects that occur in IaC scripts;}
\item{A comparison of the distribution of IaC defect categories to that of non-IaC software found in prior academic literature; and}
\item{A set of curated datasets where a mapping of defect categories and IaC scripts are provided.}
\end{itemize}

We organize the rest of the paper as following: Section~\ref{bg-rel} provides background information and prior research work relevant to our paper. Section~\ref{meth} describes our methodology conducted for this paper. We use Section~\ref{case-studies} to describe our datasets. We present our findings in Section~\ref{results}. We discuss our findings in Section~\ref{discussion}. We list the limitations of our paper in Section~\ref{limitations}. Finally, we conclude the paper in Section~\ref{conclusion}.


\section{Background and Related Work}
\label{bg-rel}

In this section, we provide background on IaC scripts and briefly describe related academic research. 

\subsection{Background}
\label{bg}

IaC is the practice of automatically defining and managing network and system configurations, and infrastructure through source code~\citep{Humble:2010:CD}. Companies widely use commercial tools such as Puppet, to implement the practice of IaC~\citep{Humble:2010:CD}~\citep{JiangAdamsMSR2015}~\citep{ShambaughRehearsal2016}. We use Puppet scripts to construct our dataset because Puppet is considered one of the most popular tools for configuration management~\citep{JiangAdamsMSR2015}~\citep{ShambaughRehearsal2016}, and has been used by companies since 2005~\citep{propuppet:book}. Typical entities of Puppet include modules and manifests~\citep{puppet-doc}. A module is a collection of manifests. Manifests are written as scripts that use a .pp extension. 

\begin{figure}[t]
\centering
\includegraphics[scale=0.80]{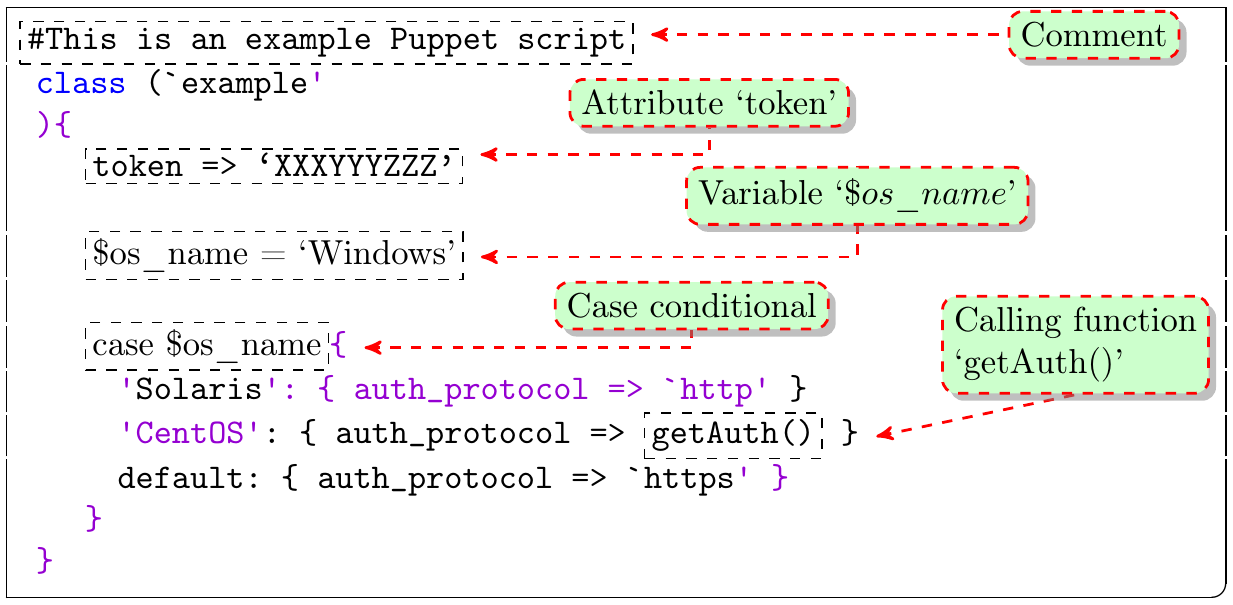}
\caption{Annotation of an example Puppet script.}
\label{fig-bg-pp}
\end{figure}

Puppet provides the utility `class' that can be used as a placeholder for the specified variables and attributes, which are used to specify configuration values. For attributes, configuration values are specified using the `$=>$' sign. For variables, configuration values are provided using the `=' sign. Similar to general purpose programming languages, code constructs such as functions/methods, comments, and conditional statements are also available for Puppet scripts. For better understanding, we provide a sample Puppet script with annotations in Figure~\ref{fig-bg-pp}.

\subsection{Related Work}
\label{related}

Our paper is related to empirical studies that have focused on IaC technologies. Sharma et al.~\citep{SharmaPuppet2016} investigated smells in IaC scripts and proposed 13 implementation and 11 design configuration smells. Hanappi et al.~\citep{Hanappi:2016:pupp:converge} investigated how convergence of Puppet scripts can be automatically tested, and proposed an automated model-based test framework. Jiang and Adams~\citep{JiangAdamsMSR2015} investigated the co-evolution of IaC scripts and other software artifacts, such as build files and source code. They reported IaC scripts to experience frequent churn. In a recent work, Rahman and Williams~\citep{me:icst2018:iac} characterized defective IaC scripts using text mining and created prediction models using text feature metrics. Bent et al.~\citep{Bent:Saner2018:Puppet} proposed and validated nine metrics to detect maintainability issues in IaC scripts. In another work, Rahman et al.~\citep{Rahman:RCOSE18} investigated the questions that practitioners ask on Stack Overflow to identify the potential challenges practitioners face while working with Puppet. We analyze the defects that occur in Puppet scripts and categorize them using the defect type attribute of ODC for the purpose of providing suggestions for process improvement. 

    
Our paper is also related to prior research studies that have categorized defects of software systems using ODC and non-ODC techniques.  Chillarege et al.~\citep{odc:original} introduced the ODC technique in 1992. They proposed eight orthogonal `defect type attributes'. Since then, researchers and practitioners have used the defect type attribute of ODC to categorize defects that occur in software. Duraes and Madeira~\citep{Duraes:Madeira2006} studied 668 faults from 12 software systems and reported that 43.4\% of the 668 defects were algorithm defects, and 21.9\% of the defects were assignment-related defects. Fonseca et al.~\citep{Fonseca2008} used ODC to categorize security defects that appear in web applications. They collected 655 security defects from six PHP web applications and reported 85.3\% of the security defects belong to the algorithm category. Zheng et al.~\citep{Zheng:Williams:2006} applied ODC on telecom-based software systems and observed an average of 35.4\% of defects belong to the algorithm category. Lutz and Mikluski~\citep{Lutz:TSE:2004} studied defect reports from seven missions of NASA and observed functional defects to be the most frequent category of 199 reported defects. Christmasson and Chillarege~\citep{Chris:1996} studied 408 IBM OS faults extracted and reported 37.7\% and 19.1\% of these faults to belong to the algorithm and assignment categories, respectively. Basso et al.~\citep{basso} studied defects from six Java-based software namely, Azureus, FreeMind, Jedit, Phex, Struts, and Tomcat, and observed the most frequent category to be algorithm defects. Cinque et al.~\citep{Cinque2014} analyzed logs from an industrial system that belong to the air traffic control system and reported that 58.9\% of the 3,159 defects were classified as algorithm defects.

The above-mentioned studies highlight the research community's interest in systematically categorizing the defects in different software systems. These studies focus on non-IaC software which highlights the lack of studies that investigate defect categories in IaC, and motivate us further to investigate defect categories of IaC scripts. Categorization of IaC-related defects can help practitioners provide actionable recommendations on how to mitigate IaC-related defects and improve the quality of their developed IaC scripts.   


\section{Categorization Methodology}
\label{meth}
In this section, we first provide definitions related to our research study and then provide the details of the methodology we used to categorize the IaC defects. 

\begin{itemize}
\item{\textbf{Defect}: An imperfection that needs to be replaced or repaired~\citep{ieee:def}.}
\item{\textbf{Defect-related commit}: A commit whose message indicates that an action was taken related to a defect.}
\item{\textbf{Defective script}: An IaC script which is listed in a defect-related commit.}
\item{\textbf{Neutral script}: An IaC script which is not listed in any defect-related commit.}
\end{itemize}

\subsection{Dataset Construction}
\label{meth-dataset}
Our methodology of dataset construction involves two steps: repository collection (Section~\ref{repo-collect}) and commit message processing (Section~\ref{commit-collect}).

\subsubsection{Repository Collection}
\label{repo-collect}
We use open source repositories to construct our datasets. An open source repository contains valuable information about the development process of an open source project, but the frequency of commits might be small indicating the repository's dormant status~\citep{MunaiahCuration2017}. This observation motivates us to apply the following selection criteria to identify repositories for mining:

\begin{itemize}
\item{\textbf{Criteria-1}: The repository must be available for download.}
\item{\textbf{Criteria-2}: At least 11\% of the files belonging to the repository must be IaC scripts. Jiang and Adams~\citep{JiangAdamsMSR2015} reported that in open source repositories IaC scripts co-exist with other types of files, such as source code and Makefiles files. They observed a median of 11\% of the files to be IaC scripts. By using a cutoff of 11\%, we collect a set of repositories that contains a sufficient amount of IaC scripts for analysis.}
\item{\textbf{Criteria-3}: The repository must have at least two commits per month. Munaiah et al.~\citep{MunaiahCuration2017} used the threshold of at least two commits per month to identify repositories which include a lower amount of development activity.}
\end{itemize} 

\subsubsection{Commit Message Processing}
\label{commit-collect}

We use two artifacts from version control systems of the selected repositories from Section~\ref{repo-collect}, to construct our datasets: (i) commits that indicate modification of IaC scripts; and (ii) issue reports that are linked with the commits. We use commits because commits contain information on how and why a file was changed. Commits can also include links to issue reports. We use issue report summaries because they can give us more insights on why IaC scripts were changed to supplement what is found in commit messages. We collect commits and other relevant information in the following manner:
\begin{itemize}
\item{First, we extract commits that were used to modify at least one IaC script. A commit lists the changes made on one or multiple files~\citep{maletic:commit:icpc2008}.}
\item{Second, we extract the message of the commit identified from the previous step. A commit includes a message, commonly referred to as a commit message. The commit messages indicate why the changes were made to the corresponding files~\citep{maletic:commit:icpc2008}.}
\item{Third, if the commit message included a unique identifier that maps the commit to an issue in the issue tracking system, we extract the identifier and use that identifier to extract the summary of the issue. We use regular expressions to extract the issue identifier. We use the corresponding issue tracking API to extract the summary of the issue; and}
\item{Fourth, we combine the commit message with any existing issue summary to construct the message for analysis. We refer to the combined message as `extended commit message (XCM)' throughout the rest of the paper. We use the extracted XCMs to categorize defects, as described in Section~\ref{categ}.} 
\end{itemize}

\subsection{Categorization of Infrastructure as Code (IaC) Script Defects}
\label{categ}

We use the defect type attribute of ODC to categorize defects. We select the ODC defect type attribute because ODC uses semantic information collected from the software system to categorize defects~\citep{odc:original}. According to the ODC defect type attribute, a defect can belong to one of the eight categories: algorithm (AL), assignment (AS), build/package/merge (B), checking (C), documentation (D), function (F), interface (I), and timing/serialization (T). 

The collected XCMs derived from commits and issue report descriptions might correspond to feature enhancement or performing maintenance tasks, which are not related to defects. As an XCM might not correspond to a defect, we added a `No defect (N)' category. An example XCM that belongs to the `No defect' category is ``add example for cobbler", as shown in Table~\ref{table-xcm-example}. Furthermore, a XCM might not to belong to any of the eight categories included in the ODC defect type attribute. Hence, we introduced the `Other (O)' category. An example XCM that belongs to the `Other' category is ``minor fixes", as shown in Table~\ref{table-xcm-example}. 

We categorize the XCMs into one of 10 categories. In the case of the eight ODC categories, we follow the criteria provided by Chillarege et al.~\citep{odc:original} and used two of our own criteria for categories `No defect', and `Other'. The criteria for each of the 10 categories are described in Table~\ref{table-odc}.  

In Table~\ref{table-odc}, the `Process Improvement Activity Suggested By ODC' column corresponds to one or more software development activities which can be improved based on the defect categories determined by ODC. For example, according to ODC, algorithm-related defects can be reduced by increasing activities that are related to coding, code inspection, unit testing, and function testing. 

\begin{table*}[]
\centering
\caption{Determining Defect Categories Based on ODC adapted from~\citep{odc:original}}
\label{table-odc}
{\footnotesize
\begin{tabular}{ p{2.8cm}  p{5.0cm}  p{2.8cm}}
\hline
\textbf{Category} & \textbf{Definition} & \textbf{Process Improvement Activity}  \\
\hline
Algorithm (AL) & Indicates efficiency or correctness problems that affect task and can be fixed by re-implementing an algorithm or local data structure. & Coding, Code Inspection, Unit Testing, Function Testing \\
\hline
Assignment (AS) & Indicates changes in a few lines of code.  & Code Inspection, Unit Testing \\
\hline
Build/Package/Merge (B) & Indicates defects due to mistakes in change management, library systems, or version control system. & Coding, Low Level Design  \\
\hline
Checking (C) & Indicates defects related to data validation and value checking. & Coding, Code Inspection, Unit Testing  \\
\hline
Documentation (D) & Indicates defects that affect publications and maintenance notes. & Coding, Publication \\
\hline
Function (F) & Indicates defects that affect significant capabilities. & Design, High Level Design Inspection, Function Testing \\
\hline
Interface (I) & Indicates defects in interacting with other components, modules, or control blocks. & Coding, System Testing \\
\hline
No Defect (N) & Indicates no defects. & Not Applicable \\
\hline
Other (O) & Indicates a defect that do not belong to the categories: AL, AS, B, C, D, F, I, N, and T. & Not Applicable \\
\hline
Timing/Serialization (T) & Indicates errors that involve real time resources and shared resources. & Low Level Design, Low Level Design Inspection, System Testing \\
\hline
\end{tabular}
}
\end{table*} 
 
We perform qualitative analysis on the collected XCMs to determine the category to which a commit belongs. We had raters with software engineering experience apply the 10 categories stated in Table~\ref{table-odc} on the collected XCMs. We record the amount of time they took to perform the categorization. After applying qualitative analysis based on the 10 defect categories, we find a mapping between each XCM and a category. We list an example XCM that belongs to each of the 10 defect categories in Table~\ref{table-xcm-example}. 


We conduct the qualitative analysis in the following manner: 

\begin{itemize}
\item{\textbf{Categorization Phase}: We randomly distribute the XCMs so that each XCM is reviewed by at least two raters to mitigate the subjectivity introduced by a single rater. Each rater determines the category of an XCM using the 10 categories presented in Table~\ref{table-odc}. We provide raters with an electronic handbook on IaC~\citep{puppet-doc}, the IEEE Anomaly Classification publication~\citep{ieee:def}, and the ODC publication~\citep{odc:original}. We do not provide any time constraint for the raters to categorize the defects. We record the agreement level amongst raters using two techniques: (a) by counting the XCMs for which the raters had the same rating; and (b) by computing the Cohen's Kappa score~\citep{cohens:kappa}.}     
\item{\textbf{Resolution Phase}: When raters disagree on the identified category, we use a resolver's opinion to resolve the disagreements. The first author of the paper is the resolver and is not involved in the categorization phase.}
\item{\textbf{Practitioner Agreement}: To evaluate the ratings in the categorization and the resolution phase, we randomly select 50 XCMs for each dataset. We contact practitioners who authored the commit message via e-mails. We ask the practitioners if they agree to our categorization of XCMs. High agreement between the raters' categorization and practitioner's feedback is an indication of how well the raters performed. The percentage of XCMs to which practitioners agreed upon are recorded and the Cohen's Kappa score is computed.}
\end{itemize}

From the qualitative analysis we determine which commits are defect-related. We use the defect-related commits to identify defective scripts. 


\begin{table*}[]
\centering
\caption{Example of Extended Commit Messages (XCMs) for Defect Categories}
\label{table-xcm-example}
\tiny{
\begin{tabular}{ p{1.75cm}  p{2.0cm}  p{2.0cm}  p{2.0cm}  p{2.0cm} }
\hline
\textbf{Category} & \textbf{Mirantis} & \textbf{Mozilla} & \textbf{Openstack} & \textbf{Wikimedia} \\
\hline
\textbf{Algorithm} & fixing deeper hash merging for firewall & bug 869897: make watch\_devices.sh logs no longer infinitely growing; my thought is logrotate.d but open to other choices here & fix middleware order of proxy pipeline and add missing modules  this patch fixes the order of the middlewares defined in the swift proxy server pipeline & nginx service should be stopped and disabled when nginx is absent \\
\hline
\textbf{Assignment} & fix syntax errors  this commit removes a couple of extraneous command introduced by copy/past errors & bug 867593 use correct regex syntax & resolved syntax error in collection & fix missing slash in puppet file url \\
\hline
\textbf{\makecell{Build/\\Package/\\Merge}} & params should have been imported; it was only working in tests by accident & bug 774638-concat::setup should depend on diffutils & fix db\_sync dependencies: this patch adds dependencies between the cinder-api and cinder-backup services to ensure that db\_sync is run before the services are up.  change-id: i7005 & fix varnish apt dependencies  these are required for the build it does. also; remove unnecessary package requires that are covered by require\_package \\
\hline
\textbf{Checking} & ensure we have iso/ directory in jenkins workspace  we share iso folder for storing fuel isos which are used for fuel-library tests & bug 1118354: ensure deploystudio user uid is $>$500 & fix check on threads & fix hadoop-hdfs-zkfc-init exec unless condition \$zookeeper\_hosts\_str was improperly set; since it was created using a non-existent local var '@zoookeeper\_hosts' \\
\hline
\textbf{Documentation} & fixed comments on pull. & bug 1253309 - followup fix to review comments & fix up doc string for workers variable change-id:ie886 & fix hadoop.pp documentation default \\
\hline
\textbf{Function} & fix for jenkins swarm slave variables & bug 1292523-puppet fails to set root password on buildduty-tools server & make class rally work  class rally is created initially by tool & fix ve restbase reverse proxy config  move the restbase domain and `v1' url routing bits into the apache config rather than the ve config. \\
\hline
\textbf{Interface} & fix for iso build broken dependencies & bug 859799-puppetagain buildbot masters won't reconfig because of missing sftp subsystem & update all missing parameters in all manifests & fix file location for interfaces \\
\hline
\textbf{No Defect} & added readme & merge bug 1178324 from default & add example for cobbler & new packages for the server \\
\hline
\textbf{Other} & fuel-stats nginx fix & summary: bug 797946: minor fixups; r=dividehex & minor fixes  & fix nginx configpackageservice ordering \\
\hline
\textbf{\makecell{Timing/\\Serialization}} & fix /etc/timezone file  add newline at end of file /etc/timezone & bug 838203-test\_alerts.html times out on ubuntu 12.04 vm & fix minimal available memory check change-id:iaad0 & fix hhvm library usage race condition  ensure that the hhvm lib `current' symlink is created before setting \/usr\/bin\/php to point to \/usr\/bin\/hhvm instead of \/usr\/bin\/php5. previously there was a potential for race conditions due to resource ordering rules \\
\hline
\end{tabular}
}
\end{table*} 


\subsection{\textbf{RQ1: How frequently do defects occur in infrastructure as code scripts?}}
\label{rq1}

We answer RQ1 by quantifying defect density and defects per month. We calculate defect density by counting defects that appear per 1000 lines (KLOC) of IaC script, similar to prior work~\citep{battin:software:2001}~\citep{mohagheghi:ICSE2004}~\citep{hatton:software1997}. We use Equation~\ref{equ-dd-kloc} to calculate defect density ($DD_{KLOC}$). $DD_{KLOC}$ gives an assessment of how frequently defects occur in IaC scripts. We select this measure of defect density, as this measure has been used as an industry standard to (i) establish a baseline for defect frequency; and (ii) helps to assess the quality of the software~\citep{harlan:cleanroom}~\citep{McConnell:CodeComplete}. 

\begin{equation}\label{equ-dd-kloc}
\begin{aligned}
DD_{KLOC} = \\ \frac{\text{number of defects for all IaC scripts}}{\bigg(\frac{\text{count of lines for all IaC scripts}}{1000}\bigg)}
\end{aligned}
\end{equation} 

Defects per month provides an overview on how frequently defects appear with the evolution of time. We compute the proportion of defects that occur every month to calculate defects per month. We use the metric `Defects per Month' and calculate this metric using Equation~\ref{equ-overall-temp-freq}: 

\begin{equation}\label{equ-overall-temp-freq}
\begin{aligned}
\text{Defects per Month ($m$)} = \\ \frac{\text{number of defects in month $m$}}{\text{total commits in month $m$}} \times 100
\end{aligned}
\end{equation}

To observe the trends for defects per month, we apply the Cox-Stuart test~\citep{CoxStuart}. The Cox Stuart test is a statistical test that compares the earlier data points to the later data points in a time series to determine whether or not the trend observant from the time series data is increasing or decreasing with statistical significance. We apply the following steps:  

\begin{itemize}[leftmargin=*]
\item{if Cox-Stuart test output shows an increasing trend with a p-value $< 0.05$, we determine the temporal trend to be `increasing'.}     
\item{if Cox-Stuart test output shows a  decreasing trend with a p-value $< 0.05$, we determine the temporal trend to be `increasing'.}        
\item{ if we cannot determine if the temporal trend as `increasing' or `decreasing', then we determine the temporal trend to be `consistent'.} 
\end{itemize}    


\subsection{\textbf{RQ2: By categorizing defects using the defect type attribute of orthogonal defect classification, what process improvement recommendations can we make for infrastructure as code development?}}
\label{rq2}

We answer RQ2 using the categorization achieved through qualitative analysis and by reporting the count of defects that belong to each defect category. We use the metric Defect Count for Category $x$ ($DCC$) calculated using Equation~\ref{equ-dmc}.

\begin{equation}\label{equ-dmc}
\begin{aligned}
\text{\small{Defect Count for Category $x$ ($DCC$)}} = \\ \bigg(\frac{\text{\small{count of defects that belong to category $x$}}}{\text{\small{total count of defects}}} \bigg) \\ \times 100 
\end{aligned}
\end{equation}

Answers to RQ2 will give us an overview on the distribution of defect categories for IaC scripts, which we use to determine what process improvement opportunities can be recommended for IaC development. In Table~\ref{table-odc} we have provided a mapping between defect categorization and the corresponding process improvement activity, as suggested by Chillarege et al.~\citep{odc:original}. 



\subsection{\textbf{RQ3: What are the differences between infrastructure as code (IaC) and non-IaC software process improvement activities, as determined by their defect category distribution reported in the literature?}}
\label{rq4}

We answer RQ3 by identifying prior research that have used the defect type attribute of ODC to categorize defects in other systems such as safety critical systems~\citep{Lutz:TSE:2004}, and operating systems~\citep{Chris:1996}. We collect necessary publications based on the following selection criteria: 
\begin{itemize}
\item{\textbf{Step-1}: The publication must cite Chillarege et al.~\citep{odc:original}'s ODC publication, be indexed by ACM Digital Library or IEEE Xplore or SpringerLink or ScienceDirect, and have been published on or after the year 2000. By selecting publications on or after the year 2000 we assume to obtain defect categories of software systems that are relevant and comparable against modern software systems such as IaC.}

\item{\textbf{Step-2}: The publication must use the defect type attribute of ODC in its original form to categorize defects of a software system. We exclude publications that cite Chillarege et al.~\citep{odc:original}'s paper as related work and not use the ODC defect type attribute for categorization. We also exclude publications that modify the ODC defect type attribute to form more defect categories, and use the modified version of ODC to categorize defects.}


\item{\textbf{Step-3}: The publication must explicitly report the software systems they studied with a distribution of defects across the ODC defect type categories and the total count of bugs/defects/faults for each software system. Along with defects, we consider bugs and faults, as in prior work researchers have used the terms bugs~\citep{Thung:Lo:ISSRE2012} and faults interchangeably with defects~\citep{pecchia:issre2012}.}

\end{itemize}

Our answer to RQ3 provides a list of software systems with distribution of defects categorized using the defect type attribute of ODC. For each software system, we report the main programming language it was built and the reference publication.  


\section{Datasets}
\label{case-studies}

We construct datasets using Puppet scripts from open source repositories maintained by four organizations: Mirantis, Mozilla, Openstack, and Wikimedia Commons. We select Puppet because it is considered as one of the most popular tools to implement IaC~\citep{JiangAdamsMSR2015}~\citep{ShambaughRehearsal2016}, and has been used by organizations since 2005~\citep{propuppet:book}. Mirantis is an organization that focuses on the development and support of cloud services such as OpenStack~\footnote{https://www.mirantis.com/}. Mozilla is an open source software community that develops, uses, and supports Mozilla products such as Mozilla Firefox~\footnote{https://www.mozilla.org/en-US/}. Openstack foundation is an open-source software platform for cloud computing where virtual servers and other resources are made available to customers~\footnote{https://www.openstack.org/}. Wikimedia Foundation is a non-profit organization that develops and distributes free educational content~\footnote{https://wikimediafoundation.org/}. 

\subsection{Repository Collection}

We apply the three selection criteria presented in Section~\ref{repo-collect} to identify the repositories that we use for analysis. We describe how many of the repositories satisfied each of the three criteria in Table~\ref{table-criteria-dataset}. Each row corresponds to the count of repositories that satisfy each criteria. For example, 26 repositories satisfy Criteria-1, for Mirantis. Altogether, we obtain 94 repositories to extract Puppet scripts from. 

\begin{table}[]
\centering
\caption{Filtering Criteria to Construct Defect Datasets}
\label{table-criteria-dataset}
{\footnotesize
\begin{tabular}{ p{1.5cm}  p{1.35cm} p{1.35cm} p{1.35cm} p{1.30cm} }
\hline
\textbf{Criteria}   & \textbf{Dataset} \\
\hline
                    & Mirantis & Mozilla & Openstack & Wikimedia \\
\hline
\textbf{Criteria-1} & 26 & 1,594 & 1,253 & 1,638 \\
\textbf{Criteria-2} & 20 & 2     & 61 & 11 \\
\textbf{Criteria-3} & 20 & 2     & 61 & 11 \\
\hline
\textbf{Final}      & 20 & 2     & 61 & 11 \\
\hline
\end{tabular}
}
\end{table}


\subsection{Commit Message Processing}

We report summary statistics on the collected repositories in Table~\ref{table-defect-dataset}. According to Table~\ref{table-defect-dataset}, for Mirantis we collect 165 Puppet scripts that map to 1,021 commits. Of these 1,021 commits, 82 commits include identifiers for bug reports. The number of lines of code for these 165 Puppet scripts is 17,564.  

\begin{table*}[]
\centering
\caption{Summary Attributes of Defect Datasets}
\label{table-defect-dataset}
{\footnotesize
\begin{tabular}{ p{2.65cm} p{1.70cm} p{1.70cm} p{1.70cm} p{1.70cm} }
\hline
\textbf{Properties}  & \textbf{Dataset} \\
                     & Mirantis & Mozilla & Openstack & Wikimedia \\
\hline
\textbf{Puppet Scripts} & 165 & 580  & 1,383 & 296 \\
\hline
\textbf{Puppet Code Size (LOC)} & 17,564 & 30,272 & 122,083 & 17,439 \\
\hline
\textbf{Defective Puppet Scripts} & 91 of 165, 55.1\% & 259 of 580, 44.6\%  & 810 of 1383, 58.5\% & 161 of 296, 54.4\% \\
\hline
\textbf{Commits with Puppet Scripts} & 1,021  & 3,074  & 7,808  & 972  \\
\hline
\textbf{Commits with Report IDs} & 82  of 1021, 8.0\% & 2764 of 3074, 89.9\% & 2252 of 7808, 28.8\% & 210 of 972, 21.6\% \\
\hline
\textbf{Defect-related Commits} &  344 of 1021, 33.7\% & 558 of 3074, 18.1\%  & 1987 of 7808, 25.4\% & 298 of 972, 30.6\% \\
\hline
\end{tabular}
}
\end{table*}


\subsection{Determining Categories of Defects}
\label{res-study-dataset}
We use 89 raters to categorize the XCMs, using the following phases:



\begin{itemize}
\item{
\textbf{Categorization Phase}: 
\begin{itemize} 
\item{\textbf{Mirantis}: We recruit students in a graduate course related to software engineering via e-mail. The number of students in the class was 58, and 32 students agreed to participate. We follow Internal Review Board (IRB) protocol, IRB\#12130, in recruitment of students and assignment of defect categorization tasks. We randomly distribute the 1,021 XCMs amongst the students such that each XCM is rated by at least two students. The average professional experience of the 32 students in software engineering is 1.9 years. On average, each student took 2.1 hours.}
\item{\textbf{Mozilla}: The second and third author of the paper, separately apply qualitative analysis on 3,074 XCMs. The second and third author, respectively, have a professional experience of three and two years in software engineering. The second and third author, respectively, took 37.0 and 51.2 hours to complete the categorization. }
\item{\textbf{Openstack}: The third and fourth author of the paper, separately, apply qualitative analysis on 7,808 XCMs from Openstack repositories. The third and fourth author, respectively, have a professional experience of two and one years in software engineering. The third and fourth author completed the categorization of the 7,808 XCMs respectively, in 80.0 and 130.0 hours.}
\item{\textbf{Wikimedia}: 54 graduate students recruited from the `Software Security' course are the raters. We randomly distribute the 972 XCMs amongst the students such that each XCM is rated by at least two students. According to our distribution, 140 XCMs are assigned to each student. The average professional experience of the 54 students in software engineering is 2.3 years. On average, each student took 2.1 hours to categorize the 140 XCMs. }
\end{itemize}
}
\item{
\textbf{Resolution Phase}: 
\begin{itemize} 
\item{\textbf{Mirantis}: Of the 1,021 XCMs, we observe agreement for 509 XCMs and disagreement for 512 XCMs, with a Cohen's Kappa score of 0.21. Based on Cohen's Kappa score, the agreement level is `fair'~\citep{Landis:Koch:Kappa:Range}.}
\item{\textbf{Mozilla}: Of the 3,074 XCMs, we observe agreement for 1,308 XCMs and disagreement for 1,766 XCMs, with a Cohen's Kappa score of 0.22. Based on Cohen's Kappa score, the agreement level is `fair'~\citep{Landis:Koch:Kappa:Range}.}
\item{\textbf{Openstack}: Of the 7,808 XCMs, we observe agreement for 3,188 XCMs, and disagreements for 4,620 XCMs. The Cohen's Kappa score was 0.21. Based on Cohen's Kappa score, the agreement level is `fair'~\citep{Landis:Koch:Kappa:Range}.}
\item{\textbf{Wikimedia}: Of the 972 XCMs, we observe agreement for 415 XCMs, and disagreements for 557 XCMs, with a Cohen's Kappa score of 0.23. Based on Cohen's Kappa score, the agreement level is `fair'~\citep{Landis:Koch:Kappa:Range}.}
\end{itemize}

The first author of the paper was the resolver, and resolved disagreements for all four datasets. In case of disagreements the resolver's categorization is considered as final.  

We observe that the raters agreement level to be `fair' for all four datasets. One possible explanation can be that the raters agreed on whether an XCM is defect-related, but disagreed on which of the 10 defect category of the defect is related to. For defect categorization, fair or poor agreement amongst raters however, is not uncommon. Henningsson et al.~\citep{Henningsson:ISESE2004} also reported a low agreement amongst raters.

\textbf{Practitioner Agreement}: We report the agreement level between the raters' and the practitioners' categorization for randomly selected 50 XCMs as following: 
\begin{itemize} 
\item{\textbf{Mirantis}: We contact three practitioners and all of them respond. We observe a 89.0\% agreement with a Cohen's Kappa score of 0.8. Based on Cohen's Kappa score, the agreement level is `substantial'~\citep{Landis:Koch:Kappa:Range}.}
\item{\textbf{Mozilla}:   We contact six practitioners and all of them respond. We observe a 94.0\% agreement with a Cohen's Kappa score of 0.9. Based on Cohen's Kappa score, the agreement level is `almost perfect'~\citep{Landis:Koch:Kappa:Range}.}
\item{\textbf{Openstack}: We contact 10 practitioners and all of them respond. We observe a 92.0\% agreement with a Cohen's Kappa score of 0.8. Based on Cohen's Kappa score, the agreement level is `substantial'~\citep{Landis:Koch:Kappa:Range}.}
\item{\textbf{Wikimedia}: We contact seven practitioners and all of them respond. We observe a 98.0\% agreement with a Cohen's Kappa score of 0.9. Based on Cohen's Kappa score, the agreement level is `almost perfect'~\citep{Landis:Koch:Kappa:Range}.}
\end{itemize}
}
\end{itemize}

We observe that the agreement between ours and the practitioners' categorization varies from 0.8 to 0.9, which is higher than that of the agreement between the raters in the Categorization Phase. One possible explanation can be related to how the resolver resolved the disagreements. The first author of the paper has industry experience in writing IaC scripts, which may help to determine categorizations that are consistent with practitioners. Another possible explanation can be related to the sample provided to the practitioners. The provided sample, even though randomly selected, may include commit messages whose categorization are relatively easy to agree upon.  


\textbf{Dataset Availability}\label{dataset-availability}: The constructed datasets used for empirical analysis are available as online~\footnote{https://doi.org/10.6084/m9.figshare.6465215}.  


\section{Results}
\label{results}
In this section, we provide empirical findings by answering the three research questions: 


\subsection{\textbf{RQ1: How frequently do defects occur in infrastructure as code scripts?}}
\label{res-rq1}

The defect density, measured in $DD_{KLOC}$  is respectively, 27.6, 18.4, 16.2, and 17.1 per 1000 LOC, respectively, for Mirantis, Mozilla, Openstack, and Wikimedia. We observe defect densities to vary from organization to another, which is consistent with prior research. For example, for a Fortran-based satellite planning software system, Basili and Perricone~\citep{Basili:Perricone:1984} have reported defect density to vary from 6.4 to 16.0, for every 1000 LOC. Mohagheghi et al.~\citep{mohagheghi:ICSE2004} have reported defect density to vary between 0.7 and 3.7 per KLOC for a telecom software system written in C, Erlang, and Java.



\begin{figure*}
\subfloat[]{
 \includegraphics[width=0.95\textwidth]{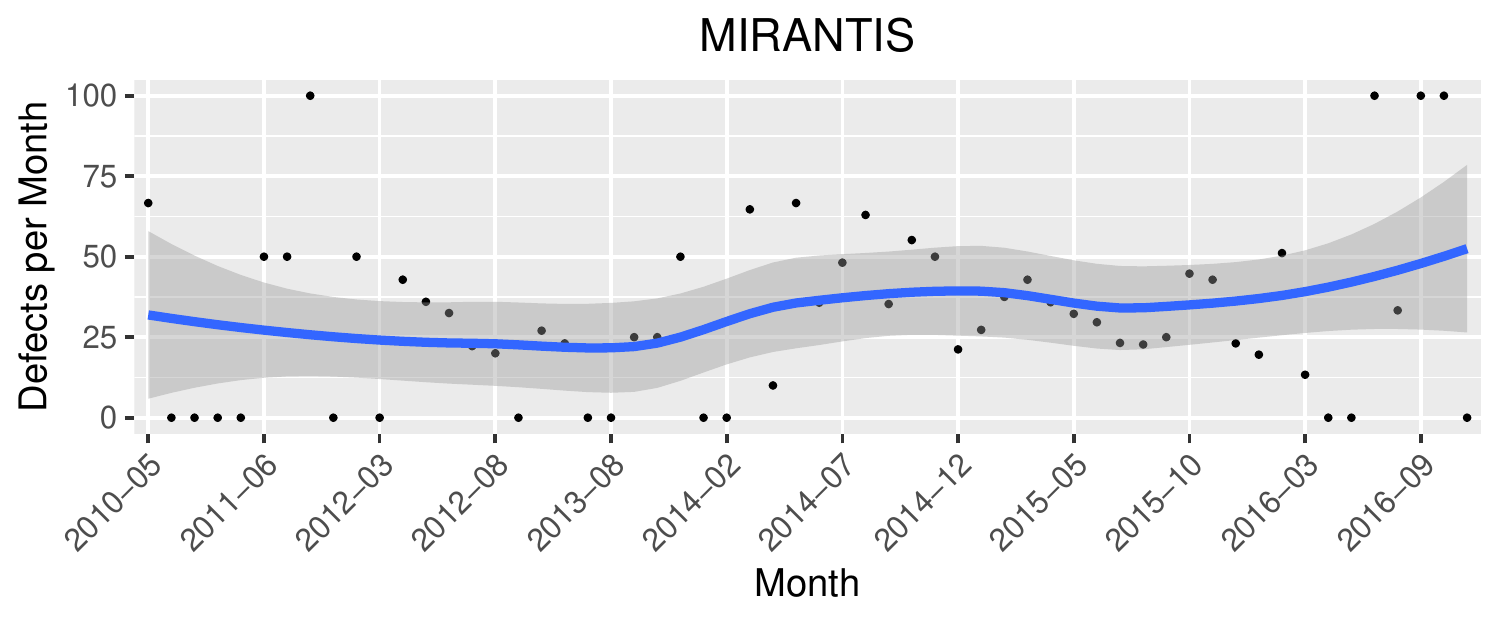}
 \label{fig-res-overall-temporal-mir}
}
\\
\subfloat[]{
 \includegraphics[width=0.95\textwidth]{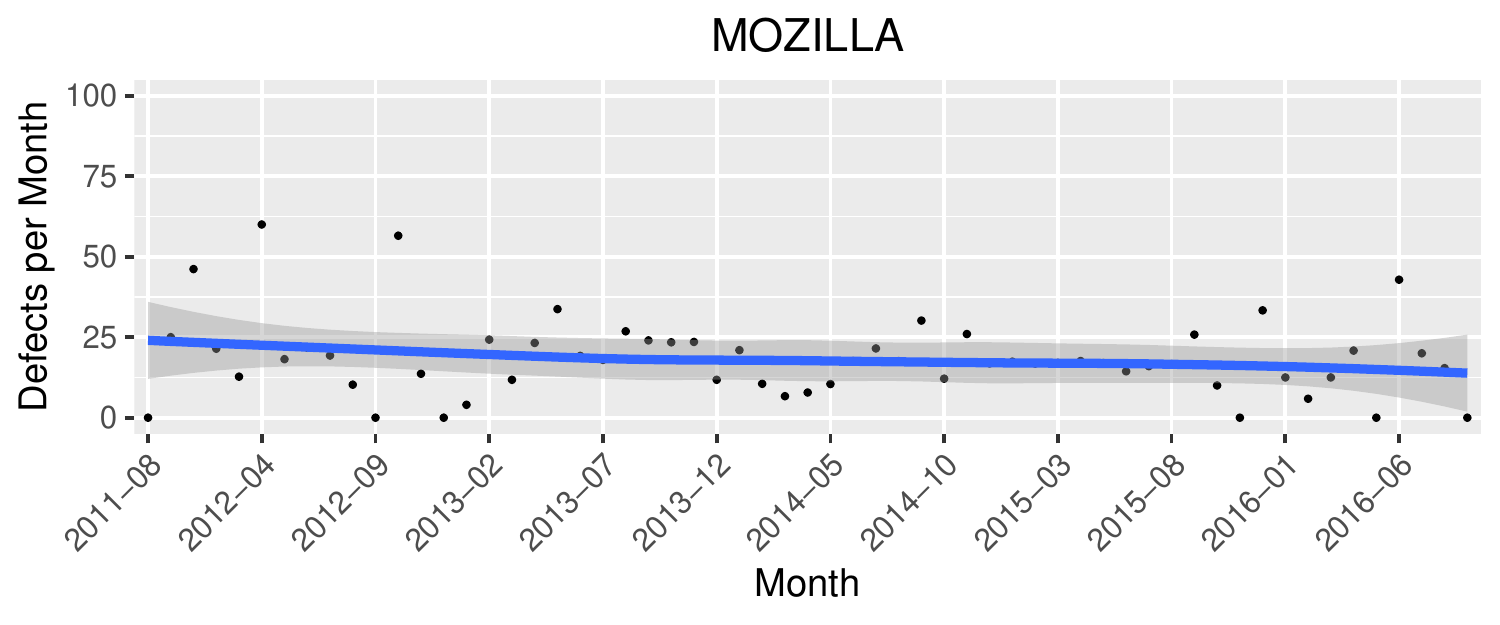}
 \label{fig-res-overall-temporal-moz}
}
\\
\subfloat[]{
 \includegraphics[width=0.95\textwidth]{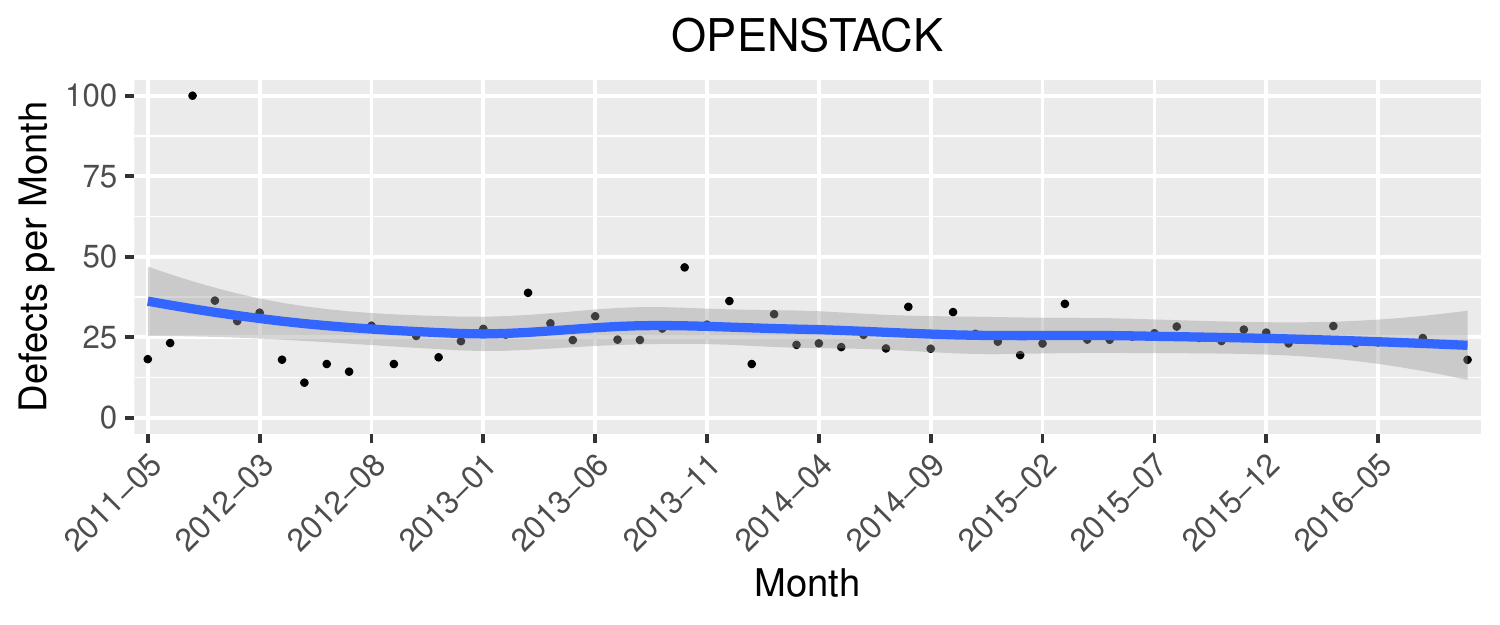}
 \label{fig-res-overall-temporal-ost}
}
\\
\subfloat[]{
 \includegraphics[width=0.95\textwidth]{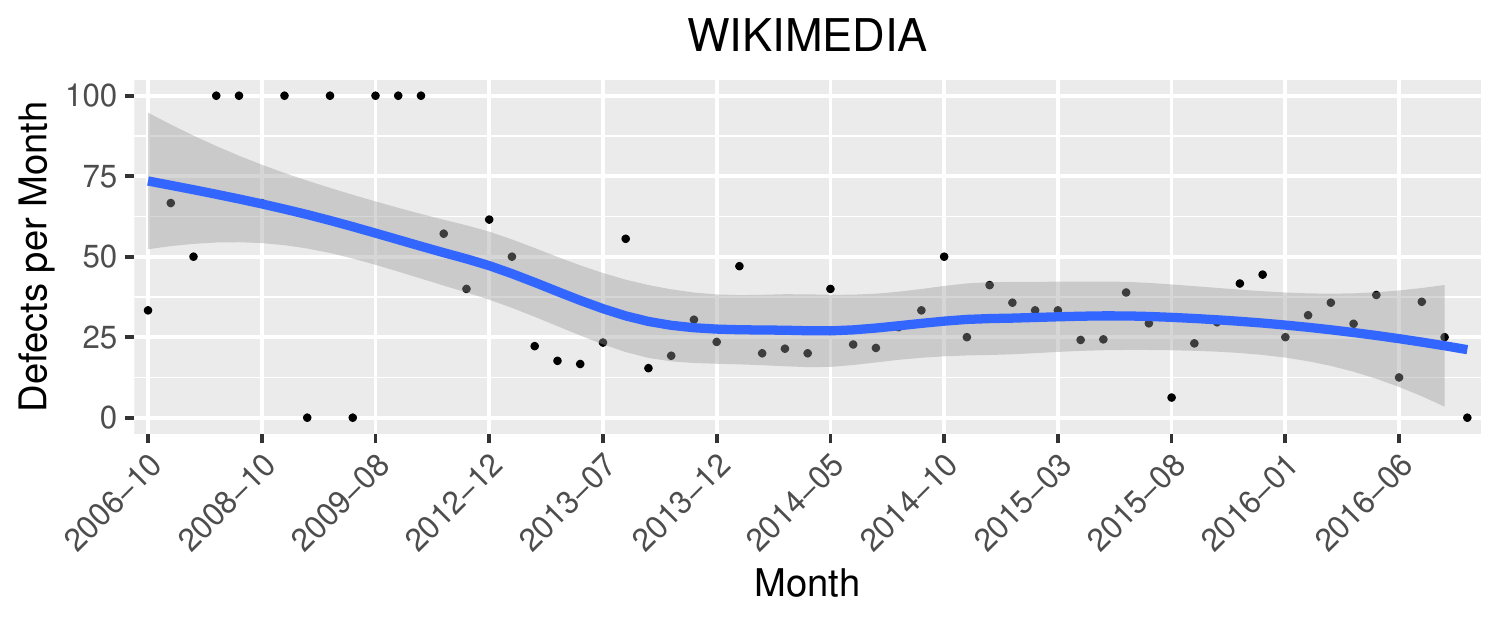}
 \label{fig-res-overall-temporal-wik}
}
\caption{Defects per month for all four datasets. Figures~\ref{fig-res-overall-temporal-mir}, ~\ref{fig-res-overall-temporal-moz}, ~\ref{fig-res-overall-temporal-ost}, and~\ref{fig-res-overall-temporal-wik} respectively presents defects per month for Mirantis, Mozilla, Openstack, and Wikimedia with smoothing. Overall, defect-related commits exhibit consistent trends over time.} 
\label{fig-res-overall-temporal} 
\end{figure*}

We report the defects per month values for the four datasets in Figure~\ref{fig-res-overall-temporal}. In Figure~\ref{fig-res-overall-temporal}, we apply smoothing to obtain visual trends. The x-axis and y-axis, respectively, presents the months and the defects per month values. According to our Cox-Stuart test results, as shown in Table~\ref{table-res-overall-temporal}, all trends are consistent. The p-values obtained from the Cox-Stuart test output for Mirantis, Mozilla, Openstack, and Wikimedia are respectively, 0.22, 0.23, 0.42, and 0.13. Our findings indicate that overall, frequency of defects do not significantly change across time for IaC scripts. 


\begin{table}
\centering
\caption{Cox-Stuart Test Results for Defects per Month}
\begin{tabular}{ c c c c c } 
\hline
\textbf{Output} & \textbf{Mirantis} & \textbf{Mozilla} & \textbf{Openstack} & \textbf{Wikimedia} \\ 
\hline
Trend           & Increasing & Decreasing & Decreasing & Decreasing \\ 
$p-value$       & 0.22 & 0.23 & 0.42 & 0.13 \\ 
\hline
\end{tabular} 
\label{table-res-overall-temporal}
\end{table}

We use our findings related to defects per month to draw parallels with hardware and non-IaC software systems. For hardware systems researchers have reported the `bathtub curve' trend, which states when initially hardware systems are put into service, the frequency of defects is high~\citep{smith1993reliability}. As time progresses, defect frequency decreases and remains constant during the `adult period'. However, after the adult period, defect frequency becomes high again, as hardware systems enter the `wear out period'~\citep{smith1993reliability}. For IaC defects we do not observe such temporal trend. 

In non-IaC software systems, initially defect frequency is high, which gradually decreases, and eventually becomes low as time progresses~\citep{hartz1996:reliability}. Eventually, all software systems enter the `obsolescence period' where defect frequency remains consistent, as no significant upgrades or changes to the software is made~\citep{hartz1996:reliability}. For Wikimedia, we observe a similar trend to that of non-IaC software systems: initially defect frequency remains high, but decreases as time progresses. However, this observation does not generalize for the other three datasets. Also, according to the Cox-Stuart test, the visible trends are not statistically validated. One possible explanation can be attributed to how organizations resolve existing defects.  For example, while fixing a certain set of defects, practitioners may be inadvertently introducing a new set of defects, which results in an overall constant trend.  

\begin{shaded}
\noindent{The defect density is 27.6, 18.4, 16.2, and 17.1 defects per KLOC respectively for Mirantis, Mozilla, Openstack, and Wikimedia. For all four datasets, we observe IaC defects to follow a consistent temporal trend.}
\end{shaded}


\subsection{\textbf{RQ2: By categorizing defects using the defect type attribute of orthogonal defect classification, what process improvement recommendations can we make for infrastructure as code development?}}
\label{res-rq2}

\begin{figure*}[htbp]
\centering
\includegraphics[scale=0.70]{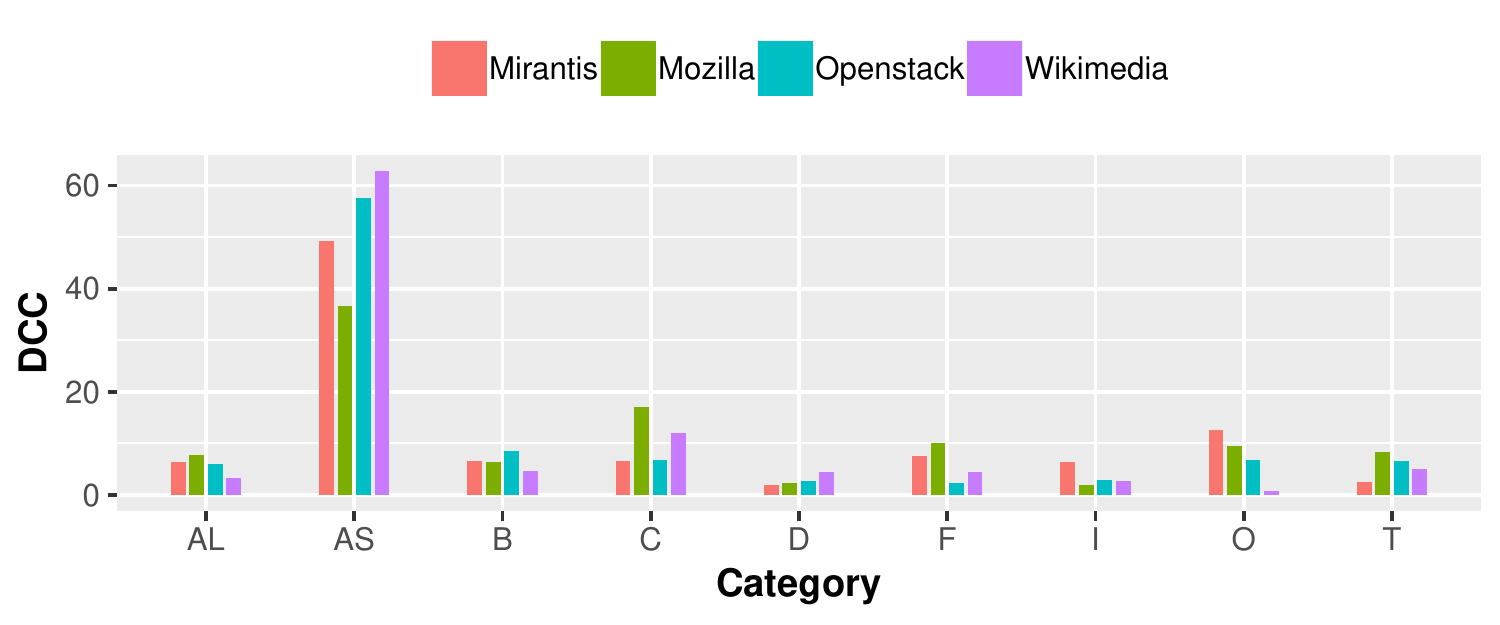}
\caption{Defect count per category (DCC) that belong to each defect category Algorithm (AL), Assignment (AS), Build/Package/Merge (B), Checking (C), Documentation (D), Function (F), Interface (I), Other (O), and Timing/Serialization (T). Defects categorized as `Assignment' is the dominant category.}
\label{fig-res-rq2-dcc}
\end{figure*}

We answer RQ2 by first presenting the values for defect count per category (DCC) that belong to each defect category mentioned in Table~\ref{table-odc}. In Figure~\ref{fig-res-rq2-dcc}, we report the DCC values for the four datasets. In Figure~\ref{fig-res-rq2-dcc}, the x-axis presents the nine defect categories, whereas, the y-axis presents DCC values for each category. We observe the dominant defect category to be assignment. As shown in Figure~\ref{fig-res-rq2-dcc}, Assignment-related defects account for 49.3\%, 36.5\%, 57.6\%, and 62.7\% of the defects, for Mirantis, Mozilla, Openstack, and Wikimedia, respectively. 

One possible explanation can be related to how practitioners utilize IaC scripts. IaC scripts are used to manage configurations and deployment infrastructure automatically~\citep{Humble:2010:CD}. For example, practitioners use IaC to provision cloud instances, such as Amazon Web Services~\citep{cito:fse2015:cloud}, or manage dependencies of software~\citep{Humble:2010:CD}. When assigning configurations of library dependencies or provisioning cloud instances practitioners might be inadvertently introducing defects. Fixing these defects involve few lines of code in IaC scripts, and these defects fall in the assignment category. Correcting syntax issues also involve a few lines of code and also belongs to the assignment category. Another possible explanation can be related to the declarative nature of Puppet scripts. Puppet provides syntax to declare and assign configurations. While assigning these configuration values practitioners may be inadvertently introducing defects.  

\begin{figure*}
\subfloat[]{
 \includegraphics[width=0.95\textwidth]{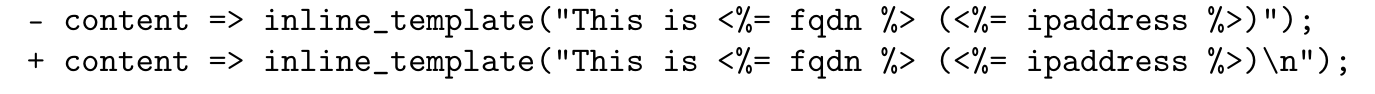}
 \label{fig-res-other-exa1}
}\\
\subfloat[]{
 \includegraphics[width=0.95\textwidth]{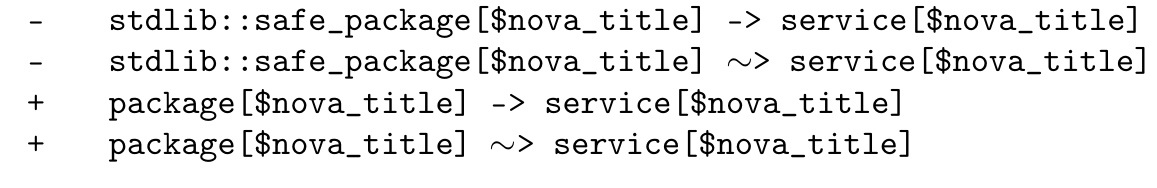}
 \label{fig-res-other-exa2}
}
\caption{Code changes in commits categorized as the `Other' category. Figures~\ref{fig-res-other-exa1} and~\ref{fig-res-other-exa2} respectively presents the code changes for two commits marked as `Other' obtained from Mozilla and Openstack.} 
\label{fig-res-other} 
\end{figure*}

As shown in Figure~\ref{fig-res-rq2-dcc}, for category Other, defect count per category is 12.5\%, 9.4\%, 6.7\%, and 0.8\%, for Mirantis, Mozilla, Openstack, and Wikimedia, respectively. This category includes XCMs that correspond to a defect but the rater was not able to identify the category of the defect. One possible explanation can be attributed to the lack of information content provided in the messages. Practitioners may not strictly adhere to the practice of writing detailed commit messages, which eventually leads to commit messages that do not provide enough information for defect categorization. For example, the commit message `minor puppetagain fixes', implies that a practitioner performed a fix-related action on an IaC script, but what category of defect was fixed remains unknown. We observe organization-based guidelines to exist on how to write better commit messages for Mozilla~\footnote{https://developer.mozilla.org/en-US/docs/Mozilla/Developer\_guide/}, Openstack~\footnote{https://wiki.openstack.org/wiki/GitCommitMessages}, and Wikimedia~\footnote{https://www.mediawiki.org/wiki/Gerrit/Commit\_message\_guidelines}. Another possible explanation can be attributed to the lack of context inherent in commit messages. The commit messages provide a summary of the changes being made, but that might not be enough to determine the defect category. Let us consider two examples in this regard, provided in Figures~\ref{fig-res-other-exa1} and~\ref{fig-res-other-exa2}. Figures~\ref{fig-res-other-exa1} and~\ref{fig-res-other-exa2} respectively presents two XCMs categorized as `Other', and obtained respectively from Mozilla and Openstack. In Figure~\ref{fig-res-other-exa2}, we observe that in the commit, a newline is being added for printing purposes, which is not captured in the commit message `summary: bug 746824: minor fixes'. From Figure~\ref{fig-res-other-exa2}, we observe that in the commit, the `stdlib::safe\_package' is being replaced by the `package' syntax, which is not captured by the corresponding commit message `minor fixes'. 

\textbf{Recommendations for IaC Development}: Our findings indicate that assignment-related defects are the most dominant category of defects. According to Chillarege et al.~\citep{odc:original}, if assignment-related defects are not discovered with code inspection and unit tests earlier, these defects can continue to grow at latter stages of development. Based on our findings, we recommend organizations to allocate more code inspection and unit testing efforts. 

\begin{shaded}
\noindent{Assignment is the most frequently occurring defect category for all four datasets: Mirantis, Mozilla, Openstack, and Wikimedia. For Mirantis, Mozilla, Openstack, and Wikimedia, respectively 49.3\%, 36.5\%, 57.6\%, and 62.7\%, of the defects belong to the category, assignment. Based on our findings, we recommend practitioners to allocate more efforts on code inspection and unit testing.}
\end{shaded}


\subsection{\textbf{RQ3: What are the differences between infrastructure as code (IaC) and non-IaC software process improvement activities, as determined by their defect category distribution reported in the literature?}}
\label{res-rq4}

We identify 26 software systems using the three steps outlined in Section~\ref{rq4}: 
\begin{itemize}
\item{\textbf{Step-1}: As of August 11, 2018, 818 publications indexed by ACM Digital Library or IEEE Xplore or SpringerLink or ScienceDirect, cited the original ODC publication~\citep{odc:original}. Of the 818 publications, 674 publications were published on or after 2000.}
\item{\textbf{Step-2}: Of these 674 publications, 16 applied the defect type attribute of ODC in its original form to categorize defects for software systems.}
\item{\textbf{Step-3}: Of these 16 publications, 7 publications explicitly mentioned the total count of defects and provided a distribution of defect categories. We use these seven publications to determine the defect categorization of 26 software systems.}
\end{itemize} 
 
In Table~\ref{table-rq4-res}, we present the categorization of defects for these 26 software systems. The `System' column reports the studied software system followed by the publication reference in which the findings were reported. The `Count' column reports the total count of defects that were studied for the software system. The `Lang.' column presents the programming language using which the system is developed. The dominant defect category for each software system is highlighted in bold.     

From Table~\ref{table-rq4-res}, we observe that for four of the 26 software systems, 40\% or more of the defect categories belonged to the assignment category. For 15 of the 26 software systems, algorithm-related defects are dominant. We also observe documentation and timing-related defects to rarely occur in the previously-studied software systems. In contrast to IaC scripts, assignment-related defects are not prevalent: assignment-related defects were the dominant category for 3 of the 26 software systems. We observe IaC scripts to have a different defect category distribution than non-IaC software system written in general purpose programming languages. 

The differences in defect category distribution may yield different set of guidelines for software process improvement. For the 15 software systems where algorithm-related defects are dominant, based on ODC, software process improvement efforts can be focused on the following activities: coding, code inspection, unit testing, and functional testing. In case of IaC scripts, as previously discussed, software process improvement efforts can be focused on code inspection and unit testing.     

\begin{shaded}
\noindent{Defects categorized as assignment are more prevalent amongst IaC scripts compared to that of non-IaC systems. Our findings suggest that process improvement activities will be different for IaC development compared to that of non-IaC software development.}
\end{shaded}

\begin{table*}[]
\captionsetup{justification=centering}
\caption{Defect Categories of Previously Studied Software Systems. Columns AL, AS, B, C, D, F, I, and T respectively correspond to the categories Algorithm, Assignment, Build/Package/Merge, Checking, Documentation, Function, Interface, and Timing/serialization.}
\label{table-rq4-res}
{\tiny
\begin{tabular}{ p{2.2cm}  p{0.5cm}  p{0.5cm}  p{0.5cm}  p{0.5cm} p{0.5cm} p{0.5cm} p{0.5cm}  p{0.5cm} p{0.5cm} p{0.5cm} p{0.5cm} p{0.5cm} }
\hline
\textbf{System} & \textbf{Lang.} & \textbf{Count} & \textbf{AL(\%)} & \textbf{AS(\%)} & \textbf{B(\%)} & \textbf{C(\%)} & \textbf{D(\%)} & \textbf{F(\%)} & \textbf{I(\%)} & \textbf{T(\%)} \\
\hline
Bourne Again Shell (BASH)~\citep{Cotroneo:TestType:JSS2013} & C & 2 & 0.0 & \textbf{100.0} & 0.0 & 0.0 & 0.0 & 0.0 & 0.0 & 0.0 \\
\hline
ZSNES-Emulator for x86~\citep{Cotroneo:TestType:JSS2013} & C, C++ &  3 & 33.3 & \textbf{66.7} & 0.0 & 0.0 & 0.0 & 0.0 & 0.0 & 0.0 \\
\hline
Pdftohtml-Pdf to html converter~\citep{Cotroneo:TestType:JSS2013} & Java &  20 & 40.0 & \textbf{55.0} & 0.0 & 5.0 & 0.0 & 0.0 & 0.0 & 0.0 \\
\hline
Firebird-Relational DBMS~\citep{Cotroneo:TestType:JSS2013} & C++ &  2 & 0.0 & \textbf{50.0} & 0.0 & \textbf{50.0} & 0.0 & 0.0 & 0.0 & 0.0 \\
\hline
Air flight application~\citep{lyu:issre2003} & C &  426 & 19.0 & 31.9 & 0.0 & 14.0 & 0.0 & \textbf{33.8} & 1.1 & 0.0 \\
\hline
Apache web server~\citep{pecchia:issre2012} & C & 1,101 & \textbf{47.6} & 26.4 & 0.0 & 12.8 & 0.0 & 0.0 & 12.9 & 0.0 \\
\hline
Joe-Tex editor~\citep{Cotroneo:TestType:JSS2013} & C &  78 & 15.3 & 25.6 & 0.0 & \textbf{44.8} & 0.0 & 0.0 & 14.1 & 0.0 \\
\hline
Middleware system for air traffic control~\citep{Cinque2014} & C & 3,159 & \textbf{58.9} & 24.5 & 0.0 & 1.7 & 0.0 & 0.0 & 14.8 & 0.0 \\
\hline
ScummVM-Interpreter for adventure engines~\citep{Cotroneo:TestType:JSS2013} & C++ &  74 & \textbf{56.7} & 24.3 & 0.0 & 8.1 & 0.0 & 6.7 & 4.0 & 0.0 \\
\hline
Linux kernel~\citep{Cotroneo:TestType:JSS2013} & C &  93 & \textbf{33.3} & 22.5 & 0.0 & 25.8 & 0.0 & 12.9 & 5.3 & 0.0 \\
\hline
Vim-Linux editor~\citep{Cotroneo:TestType:JSS2013} & C &  249 & \textbf{44.5} & 21.2 & 0.0 & 22.4 & 0.0 & 5.2 & 6.4 & 0.0 \\
\hline
MySQL DBMS~\citep{pecchia:issre2012} & C, C++ &  15,102 & \textbf{52.9} & 20.5 & 0.0 & 15.3 & 0.0 & 0.0 & 11.2 & 0.0 \\
\hline
CDEX-CD digital audio data extractor~\citep{Cotroneo:TestType:JSS2013} & C, C++, Python &  11 & 9.0 & 18.1 & 0.0 & 18.1 & 0.0 & 0.0 & \textbf{54.5} & 0.0 \\
\hline
Struts~\citep{basso} & Java &  99 & \textbf{48.4} & 18.1 & 0.0 & 9.0 & 0.0 & 4.0 & 20.2 & 0.0 \\
\hline
Safety critical system for NASA spacecraft~\citep{Lutz:TSE:2004} & Java & 199 & 17.0 & 15.5 & 2.0 & 0.0 & 2.5 & \textbf{29.1} & 5.0 & 10.5 \\
\hline
Azureus~\citep{basso} & Java & 125 & \textbf{36.8} & 15.2 & 0.0 & 11.2 & 0.0 & 28.8 & 8.0 & 0.0 \\
\hline
Phex~\citep{basso} & Java & 20 & \textbf{60.0} & 15.0 & 0.0 & 5.0 & 0.0 & 10.0 & 10.0 & 0.0 \\
\hline
TAO Open DDS~\citep{pecchia:issre2012} & Java & 1,184 & \textbf{61.4} & 14.4 & 0.0 & 11.7 & 0.0 & 0.0 & 12.4 & 0.0 \\
\hline
JEdit~\citep{basso} & Java & 71 & \textbf{36.6} & 14.0 & 0.0 & 11.2 & 0.0 & 25.3 & 12.6 & 0.0 \\
\hline
Tomcat~\citep{basso} & Java & 169 & \textbf{57.9} & 12.4 & 0.0 & 13.6 & 0.0 & 2.3 & 13.6 & 0.0 \\
\hline
FreeCvi-Strategy game~\citep{Cotroneo:TestType:JSS2013} & Java & 53 & \textbf{52.8} & 11.3 & 0.0 & 13.2 & 0.0 & 15.0 & 7.5 & 0.0 \\
\hline
FreeMind~\citep{basso} & Java & 90 & \textbf{46.6} & 11.1 & 0.0 & 2.2 & 0.0 & 28.8 & 11.1 & 0.0 \\
\hline
MinGW-Minimalist GNU for Windows~\citep{Cotroneo:TestType:JSS2013} & C & 60 & \textbf{46.6} & 10.0 & 0.0 & 38.3 & 0.0 & 0.0 & 5.0 & 0.0 \\
\hline
Java Enterprise Framework~\citep{Gupta:EMSE2009} & Java & 223 & 3.1 & 16.7 & \textbf{36.0} & 3.1 & 0.0 & 21.2 & 19.9 & 0.0 \\
\hline
Digital Cargo Files~\citep{Gupta:EMSE2009} & Java & 438 & 3.3 & 12.4 & 31.0 & 10.4 & 0.5 & \textbf{31.3} & 9.9 & 1.2 \\
\hline
Shipment and Allocation~\citep{Gupta:EMSE2009} & Java & 649 & 22.8 & 6.7 & 18.7 & 10.9 & 0.5 & \textbf{29.8} & 9.3 & 1.6 \\
\hline
\hline
Mirantis [This paper] & Puppet & 344 & 6.5 & \textbf{49.3} & 6.7 & 1.9 & 7.5 & 6.4 & 12.5 & 2.6 \\
\hline
Mozilla [This paper] & Puppet & 558 & 7.7 & \textbf{36.5} & 6.4 & 17.0 & 2.3 & 10.0 & 1.9 & 8.4 \\
\hline
Openstack [This paper] & Puppet & 1987 & 5.9 & \textbf{57.5} & 8.6 & 6.7 & 2.6 & 2.4 & 2.9 & 6.5 \\
\hline
Wikimedia [This paper] & Puppet & 298 & 3.3 & \textbf{62.7} & 4.7 & 12.0 & 4.3 & 4.3 & 2.6 & 5.1 \\
\hline
\hline
\end{tabular}
}
\end{table*}


\section{Discussion}
\label{discussion}
In this section, we discuss our findings with possible implications:

\subsection{Implications for Process Improvement} 
One finding our paper is the prevalence of assignment-related defects in IaC scripts. Software teams can use this finding to improve their process in two possible ways: first, they can use the practice of code review for developing IaC scripts. Code reviews can be conducted using automated tools and/or team members' manual reviews. For example, through code reviews software teams can pinpoint the correct value of configurations at development stage. Automated code review tools such as linters can also help in detecting and fixing syntax issues of IaC scripts at the development stage. Typically, IaC scripts are used by organizations that have implemented CD, and for these organizations, Kim et al.~\citep{GeneKim:DevOps2016} recommends manual peer review methods such as pair programming to improve code quality.    

Second, software teams might benefit from unit testing of IaC scripts to reduce defects related to configuration assignment. We have observed from Section~\ref{res-rq2} that IaC-related defects are mostly related to the assignment category which includes improper assignment of configuration values and syntax errors. Practitioners can test if correct configuration values are assigned by writing unit tests for components of IaC scripts. In this manner, instead of catching defects at run-time that might lead to real-world consequences e.g. the problem reported by Wikimedia Commons~\footnote{https://wikitech.wikimedia.org/wiki/Incident documentation/20170118-Labs}, software teams might be able to catch defects in IaC scripts at the development stages with the help of testing. 

\subsection{Future Research}
Our findings have the potential to facilitate further research in the area of IaC defects. In Section~\ref{results} we have observed the process differences that occur between organizations, and future research can systematically investigate if there are process differences in IaC development and why they exist. We have applied a qualitative process to categorize defects using the defect type attribute of ODC. We acknowledge that our process is manual and labor-intensive. We advocate for future research that can look into how the process of ODC can be automated. In future researchers can also investigate why the observed defects occur, and provide a causal analysis of IaC defects. 


\section{Threats To Validity}
\label{limitations}
We describe the threats to validity of our paper as following: 

\begin{itemize}

\item{\textbf{Conclusion Validity}: Our approach is based on qualitative analysis, where raters categorized XCMs, and assigned defect categories. We acknowledge that the process is susceptible human judgment, and the raters' experience can bias the categories assigned. The accompanying human subjectivity can influence the distribution of the defect category for IaC scripts of interest. We mitigated this threat by assigning multiple raters for the same set of XCMs. Next, we used a resolver, who resolved the disagreements. Further, we cross-checked our categorization with practitioners who authored the XCMs, and observed `substantial' to `almost perfect' agreement. 
}

\item{\textbf{Internal Validity}: We have used a combination of commit messages and issue report descriptions to determine if an IaC script is associated with a defect. We acknowledge that these messages might not have given the full context for the raters. Other sources of information such as practitioner input, and code changes that take place in each commit could have provided the raters better context to categorize the XCMs. 

In our paper, we have used a defect-related commit message as a defect. We acknowledge that in a defect-related commit message more than one defects can be fixed, but may not be expressed in the commit message and/or accompanied issue report description. 

We acknowledge that we have not used the trigger attribute of ODC, as our data sources do not include any information from which we can infer trigger attributes of ODC such as, `design conformance', `side effects', and `concurrency'.
}

\item{\textbf{Construct validity}: Our process of using human raters to determine defect categories can be limiting, as the process is susceptible to mono-method bias, where subjective judgment of raters can influence the findings. We mitigated this threat by using multiple raters.  

Also, for Mirantis and Wikimedia, we used graduate students who performed the categorization as part of their class work. Students who participated in the categorization process can be subject to evaluation apprehension, i.e. consciously or sub-consciously relating their performance with the grades they would achieve for the course. We mitigated this threat by clearly explaining to the students that their performance in the categorization process would not affect their grades. 

The raters involved in the categorization process had professional experience in software engineering for at two years on average. Their experience in software engineering may make the raters curious about the expected outcomes of the categorization process, which may effect the distribution of the categorization process. Furthermore, the resolver also has professional experience in software engineering and IaC script development, which could influence the outcome of the defect category distribution. 
}

\item{\textbf{External Validity}: Our scripts are collected from the OSS domain, and not from proprietary sources. Our findings are subject to external validity, as our findings may not be generalizable.

We construct our datasets using Puppet, which is a declarative language. Our findings may not generalize for IaC scripts that use an imperative form of language. We hope to mitigate this limitation by analyzing IaC scripts written using imperative form of languages.  
} 

\end{itemize}



\section{Conclusion}
\label{conclusion}
Use of IaC scripts is crucial for the automated maintenance of software delivery and deployment infrastructure. Similar to software source code, IaC scripts churn frequently, and can include defects which can have serious consequences. IaC defect categorization can help organizations to identify opportunities to improve IaC development. We have conducted an empirical analysis using four datasets from four organizations namely, Mirantis, Mozilla, Openstack, and Wikimedia Commons. With 89 raters, we apply the defect type attribute of the orthogonal defect classification (ODC) methodology to categorize the defects. For all four datasets, we have observed that majority of the defects are related to assignment i.e. defects related to syntax errors and erroneous configuration assignments. We have observed compared to IaC scripts, assignment-related defects occur less frequently in non-IaC software. We also have observed the defect density is 27.6, 18.4, 16.2, and 17.1 defects per KLOC respectively for Mirantis, Mozilla, Openstack, and Wikimedia. For all four datasets, we observe IaC defects to follow a consistent temporal trend. We hope our findings will facilitate further research in IaC defect analysis.


\begin{acknowledgements}
The Science of Security Lablet at the North Carolina State University supported this research study. We thank the Realsearch research group members for their useful feedback. We also thank the practitioners who answered our questions.
\end{acknowledgements}

\bibliographystyle{spbasic}      

\bibliography{emse}                

\begin{thebibliography}{45}
\providecommand{\natexlab}[1]{#1}
\providecommand{\url}[1]{{#1}}
\providecommand{\urlprefix}{URL }
\expandafter\ifx\csname urlstyle\endcsname\relax
  \providecommand{\doi}[1]{DOI~\discretionary{}{}{}#1}\else
  \providecommand{\doi}{DOI~\discretionary{}{}{}\begingroup
  \urlstyle{rm}\Url}\fi
\providecommand{\eprint}[2][]{\url{#2}}

\bibitem[{Alali et~al(2008)Alali, Kagdi, and Maletic}]{maletic:commit:icpc2008}
Alali A, Kagdi H, Maletic JI (2008) What's a typical commit? a characterization
  of open source software repositories. In: 2008 16th IEEE International
  Conference on Program Comprehension, pp 182--191, \doi{10.1109/ICPC.2008.24}

\bibitem[{Basili and Perricone(1984)}]{Basili:Perricone:1984}
Basili VR, Perricone BT (1984) Software errors and complexity: An empirical
  investigation0. Commun ACM 27(1):42--52, \doi{10.1145/69605.2085},
  \urlprefix\url{http://doi.acm.org/10.1145/69605.2085}

\bibitem[{Basso et~al(2009)Basso, Moraes, Sanches, and Jino}]{basso}
Basso T, Moraes R, Sanches BP, Jino M (2009) An investigation of java faults
  operators derived from a field data study on java software faults. In:
  Workshop de Testes e Toler{\^a}ncia a Falhas, pp 1--13

\bibitem[{Battin et~al(2001)Battin, Crocker, Kreidler, and
  Subramanian}]{battin:software:2001}
Battin RD, Crocker R, Kreidler J, Subramanian K (2001) Leveraging resources in
  global software development. IEEE Software 18(2):70--77,
  \doi{10.1109/52.914750}

\bibitem[{van~der Bent et~al(2018)van~der Bent, Hage, Visser, and
  Gousios}]{Bent:Saner2018:Puppet}
van~der Bent E, Hage J, Visser J, Gousios G (2018) How good is your puppet? an
  empirically defined and validated quality model for puppet. In: 2018 IEEE
  25th International Conference on Software Analysis, Evolution and
  Reengineering (SANER), pp 164--174, \doi{10.1109/SANER.2018.8330206}

\bibitem[{Butcher et~al(2002)Butcher, Munro, and Kratschmer}]{butcher:ibm:odc}
Butcher M, Munro H, Kratschmer T (2002) Improving software testing via odc:
  Three case studies. IBM Systems Journal 41(1):31--44,
  \doi{10.1147/sj.411.0031}

\bibitem[{Chillarege et~al(1992)Chillarege, Bhandari, Chaar, Halliday, Moebus,
  Ray, and Wong}]{odc:original}
Chillarege R, Bhandari IS, Chaar JK, Halliday MJ, Moebus DS, Ray BK, Wong MY
  (1992) Orthogonal defect classification-a concept for in-process
  measurements. IEEE Transactions on Software Engineering 18(11):943--956,
  \doi{10.1109/32.177364}

\bibitem[{Christmansson and Chillarege(1996)}]{Chris:1996}
Christmansson J, Chillarege R (1996) Generation of an error set that emulates
  software faults based on field data. In: Proceedings of Annual Symposium on
  Fault Tolerant Computing, pp 304--313, \doi{10.1109/FTCS.1996.534615}

\bibitem[{Cinque et~al(2014)Cinque, Cotroneo, Corte, and Pecchia}]{Cinque2014}
Cinque M, Cotroneo D, Corte RD, Pecchia A (2014) Assessing direct monitoring
  techniques to analyze failures of critical industrial systems. In: 2014 IEEE
  25th International Symposium on Software Reliability Engineering, pp
  212--222, \doi{10.1109/ISSRE.2014.30}

\bibitem[{Cito et~al(2015)Cito, Leitner, Fritz, and Gall}]{cito:fse2015:cloud}
Cito J, Leitner P, Fritz T, Gall HC (2015) The making of cloud applications: An
  empirical study on software development for the cloud. In: Proceedings of the
  2015 10th Joint Meeting on Foundations of Software Engineering, ACM, New
  York, NY, USA, ESEC/FSE 2015, pp 393--403, \doi{10.1145/2786805.2786826},
  \urlprefix\url{http://doi.acm.org/10.1145/2786805.2786826}

\bibitem[{Cohen(1960)}]{cohens:kappa}
Cohen J (1960) A coefficient of agreement for nominal scales. Educational and
  Psychological Measurement 20(1):37--46, \doi{10.1177/001316446002000104},
  \urlprefix\url{http://dx.doi.org/10.1177/001316446002000104},
  \eprint{http://dx.doi.org/10.1177/001316446002000104}

\bibitem[{Cotroneo et~al(2013)Cotroneo, Pietrantuono, and
  Russo}]{Cotroneo:TestType:JSS2013}
Cotroneo D, Pietrantuono R, Russo S (2013) Testing techniques selection based
  on odc fault types and software metrics. J Syst Softw 86(6):1613--1637,
  \doi{10.1016/j.jss.2013.02.020},
  \urlprefix\url{http://dx.doi.org/10.1016/j.jss.2013.02.020}

\bibitem[{Cox and Stuart(1955)}]{CoxStuart}
Cox DR, Stuart A (1955) Some quick sign tests for trend in location and
  dispersion. Biometrika 42(1/2):80--95,
  \urlprefix\url{http://www.jstor.org/stable/2333424}

\bibitem[{Duraes and Madeira(2006)}]{Duraes:Madeira2006}
Duraes JA, Madeira HS (2006) Emulation of software faults: A field data study
  and a practical approach. IEEE Trans Softw Eng 32(11):849--867,
  \doi{10.1109/TSE.2006.113},
  \urlprefix\url{http://dx.doi.org/10.1109/TSE.2006.113}

\bibitem[{Fonseca and Vieira(2008)}]{Fonseca2008}
Fonseca J, Vieira M (2008) Mapping software faults with web security
  vulnerabilities. In: 2008 IEEE International Conference on Dependable Systems
  and Networks With FTCS and DCC (DSN), pp 257--266,
  \doi{10.1109/DSN.2008.4630094}

\bibitem[{Gupta et~al(2009)Gupta, Li, Conradi, R{\o}nneberg, and
  Landre}]{Gupta:EMSE2009}
Gupta A, Li J, Conradi R, R{\o}nneberg H, Landre E (2009) A case study
  comparing defect profiles of a reused framework and of applications reusing
  it. Empirical Softw Engg 14(2):227--255, \doi{10.1007/s10664-008-9081-9},
  \urlprefix\url{http://dx.doi.org/10.1007/s10664-008-9081-9}

\bibitem[{Hanappi et~al(2016)Hanappi, Hummer, and
  Dustdar}]{Hanappi:2016:pupp:converge}
Hanappi O, Hummer W, Dustdar S (2016) Asserting reliable convergence for
  configuration management scripts. SIGPLAN Not 51(10):328--343,
  \doi{10.1145/3022671.2984000},
  \urlprefix\url{http://doi.acm.org/10.1145/3022671.2984000}

\bibitem[{Harlan(1987)}]{harlan:cleanroom}
Harlan D (1987) Cleanroom software engineering

\bibitem[{Hartz et~al(1996)Hartz, Walker, and Mahar}]{hartz1996:reliability}
Hartz MA, Walker EL, Mahar D (1996) Introduction to software reliability: A
  state of the art review. The Center

\bibitem[{Hatton(1997)}]{hatton:software1997}
Hatton L (1997) Reexamining the fault density-component size connection. IEEE
  Softw 14(2):89--97, \doi{10.1109/52.582978},
  \urlprefix\url{http://dx.doi.org/10.1109/52.582978}

\bibitem[{Henningsson and Wohlin(2004)}]{Henningsson:ISESE2004}
Henningsson K, Wohlin C (2004) Assuring fault classification agreement " an
  empirical evaluation. In: Proceedings of the 2004 International Symposium on
  Empirical Software Engineering, IEEE Computer Society, Washington, DC, USA,
  ISESE '04, pp 95--104, \doi{10.1109/ISESE.2004.13},
  \urlprefix\url{http://dx.doi.org/10.1109/ISESE.2004.13}

\bibitem[{Humble and Farley(2010)}]{Humble:2010:CD}
Humble J, Farley D (2010) Continuous Delivery: Reliable Software Releases
  Through Build, Test, and Deployment Automation, 1st edn. Addison-Wesley
  Professional

\bibitem[{IEEE(2010)}]{ieee:def}
IEEE (2010) Ieee standard classification for software anomalies. IEEE Std
  1044-2009 (Revision of IEEE Std 1044-1993) pp 1--23,
  \doi{10.1109/IEEESTD.2010.5399061}

\bibitem[{Jiang and Adams(2015)}]{JiangAdamsMSR2015}
Jiang Y, Adams B (2015) Co-evolution of infrastructure and source code: An
  empirical study. In: Proceedings of the 12th Working Conference on Mining
  Software Repositories, IEEE Press, Piscataway, NJ, USA, MSR '15, pp 45--55,
  \urlprefix\url{http://dl.acm.org/citation.cfm?id=2820518.2820527}

\bibitem[{Kim et~al(2016)Kim, Debois, Willis, and Humble}]{GeneKim:DevOps2016}
Kim G, Debois P, Willis J, Humble J (2016) The DevOps Handbook: How to Create
  World-Class Agility, Reliability, and Security in Technology Organizations.
  IT Revolution Press

\bibitem[{Labs(2017)}]{puppet-doc}
Labs P (2017) {Puppet Documentation}. \url{https://docs.puppet.com/}, [Online;
  accessed 19-July-2017]

\bibitem[{Landis and Koch(1977)}]{Landis:Koch:Kappa:Range}
Landis JR, Koch GG (1977) The measurement of observer agreement for categorical
  data. Biometrics 33(1):159--174,
  \urlprefix\url{http://www.jstor.org/stable/2529310}

\bibitem[{Levy and Chillarege(2003)}]{comverse:odc:ram}
Levy D, Chillarege R (2003) Early warning of failures through alarm analysis a
  case study in telecom voice mail systems. In: 14th International Symposium on
  Software Reliability Engineering, 2003. ISSRE 2003., pp 271--280,
  \doi{10.1109/ISSRE.2003.1251049}

\bibitem[{Lutz and Mikulski(2004)}]{Lutz:TSE:2004}
Lutz RR, Mikulski IC (2004) Empirical analysis of safety-critical anomalies
  during operations. IEEE Transactions on Software Engineering 30(3):172--180,
  \doi{10.1109/TSE.2004.1271171}

\bibitem[{Lyu et~al(2003)Lyu, Huang, Sze, and Cai}]{lyu:issre2003}
Lyu MR, Huang Z, Sze SKS, Cai X (2003) An empirical study on testing and fault
  tolerance for software reliability engineering. In: 14th International
  Symposium on Software Reliability Engineering, 2003. ISSRE 2003., pp
  119--130, \doi{10.1109/ISSRE.2003.1251036}

\bibitem[{McConnell(2004)}]{McConnell:CodeComplete}
McConnell S (2004) Code complete - a practical handbook of software
  construction, 2nd edition

\bibitem[{McCune and Jeffrey(2011)}]{propuppet:book}
McCune JT, Jeffrey (2011) Pro Puppet, 1st edn. Apress,
  \doi{10.1007/978-1-4302-3058-8},
  \urlprefix\url{https://www.springer.com/gp/book/9781430230571}

\bibitem[{Mohagheghi et~al(2004)Mohagheghi, Conradi, Killi, and
  Schwarz}]{mohagheghi:ICSE2004}
Mohagheghi P, Conradi R, Killi OM, Schwarz H (2004) An empirical study of
  software reuse vs. defect-density and stability. In: Proceedings of the 26th
  International Conference on Software Engineering, IEEE Computer Society,
  Washington, DC, USA, ICSE '04, pp 282--292,
  \urlprefix\url{http://dl.acm.org/citation.cfm?id=998675.999433}

\bibitem[{Munaiah et~al(2017)Munaiah, Kroh, Cabrey, and
  Nagappan}]{MunaiahCuration2017}
Munaiah N, Kroh S, Cabrey C, Nagappan M (2017) Curating github for engineered
  software projects. Empirical Software Engineering pp 1--35,
  \doi{10.1007/s10664-017-9512-6},
  \urlprefix\url{http://dx.doi.org/10.1007/s10664-017-9512-6}

\bibitem[{Parnin et~al(2017)Parnin, Helms, Atlee, Boughton, Ghattas, Glover,
  Holman, Micco, Murphy, Savor, Stumm, Whitaker, and Williams}]{parnin:adages}
Parnin C, Helms E, Atlee C, Boughton H, Ghattas M, Glover A, Holman J, Micco J,
  Murphy B, Savor T, Stumm M, Whitaker S, Williams L (2017) The top 10 adages
  in continuous deployment. IEEE Software 34(3):86--95,
  \doi{10.1109/MS.2017.86}

\bibitem[{Pecchia and Russo(2012)}]{pecchia:issre2012}
Pecchia A, Russo S (2012) Detection of software failures through event logs: An
  experimental study. In: 2012 IEEE 23rd International Symposium on Software
  Reliability Engineering, pp 31--40, \doi{10.1109/ISSRE.2012.24}

\bibitem[{Puppet(2018)}]{ambit:pup}
Puppet (2018) Ambit energy's competitive advantage? it's really a devops
  software company. Tech. rep., Puppet,
  \urlprefix\url{https://puppet.com/resources/case-study/ambit-energy}

\bibitem[{Rahman and Williams(2018)}]{me:icst2018:iac}
Rahman A, Williams L (2018) Characterizing defective configuration scripts used
  for continuous deployment. In: 2018 IEEE 11th International Conference on
  Software Testing, Verification and Validation (ICST), pp 34--45,
  \doi{10.1109/ICST.2018.00014}

\bibitem[{Rahman et~al(2018)Rahman, Partho, Morrison, and
  Williams}]{Rahman:RCOSE18}
Rahman A, Partho A, Morrison P, Williams L (2018) What questions do programmers
  ask about configuration as code? In: Proceedings of the 4th International
  Workshop on Rapid Continuous Software Engineering, ACM, New York, NY, USA,
  RCoSE '18, pp 16--22, \doi{10.1145/3194760.3194769},
  \urlprefix\url{http://doi.acm.org/10.1145/3194760.3194769}

\bibitem[{Rahman et~al(2015)Rahman, Helms, Williams, and
  Parnin}]{me:agile:cd2015}
Rahman AAU, Helms E, Williams L, Parnin C (2015) Synthesizing continuous
  deployment practices used in software development. In: Proceedings of the
  2015 Agile Conference, IEEE Computer Society, Washington, DC, USA, AGILE '15,
  pp 1--10, \doi{10.1109/Agile.2015.12},
  \urlprefix\url{http://dx.doi.org/10.1109/Agile.2015.12}

\bibitem[{Shambaugh et~al(2016)Shambaugh, Weiss, and
  Guha}]{ShambaughRehearsal2016}
Shambaugh R, Weiss A, Guha A (2016) Rehearsal: A configuration verification
  tool for puppet. SIGPLAN Not 51(6):416--430, \doi{10.1145/2980983.2908083},
  \urlprefix\url{http://doi.acm.org/10.1145/2980983.2908083}

\bibitem[{Sharma et~al(2016)Sharma, Fragkoulis, and
  Spinellis}]{SharmaPuppet2016}
Sharma T, Fragkoulis M, Spinellis D (2016) Does your configuration code smell?
  In: Proceedings of the 13th International Conference on Mining Software
  Repositories, ACM, New York, NY, USA, MSR '16, pp 189--200,
  \doi{10.1145/2901739.2901761},
  \urlprefix\url{http://doi.acm.org/10.1145/2901739.2901761}

\bibitem[{Smith(1993)}]{smith1993reliability}
Smith AM (1993) Reliability-centered maintenance, vol~83. McGraw-Hill New York

\bibitem[{Thung et~al(2012)Thung, Wang, Lo, and Jiang}]{Thung:Lo:ISSRE2012}
Thung F, Wang S, Lo D, Jiang L (2012) An empirical study of bugs in machine
  learning systems. In: 2012 IEEE 23rd International Symposium on Software
  Reliability Engineering, pp 271--280, \doi{10.1109/ISSRE.2012.22}

\bibitem[{Zheng et~al(2006)Zheng, Williams, Nagappan, Snipes, Hudepohl, and
  Vouk}]{Zheng:Williams:2006}
Zheng J, Williams L, Nagappan N, Snipes W, Hudepohl JP, Vouk MA (2006) On the
  value of static analysis for fault detection in software. IEEE Transactions
  on Software Engineering 32(4):240--253, \doi{10.1109/TSE.2006.38}

\end{thebibliography}

\end{document}